\documentclass[12pt]{article}
\usepackage{mathptmx}       
\usepackage{helvet}         
\usepackage{courier}        
\usepackage{type1cm}        
\usepackage{amsmath}
\usepackage[english]{babel}
   
\usepackage[margin=.7in]{geometry}     

\usepackage{makeidx}         
\usepackage{graphicx}        
\usepackage{multicol}        
\usepackage[bottom]{footmisc}

\newcounter{nq}

\usepackage{cite,url,ifthen,multicol,amssymb,lmodern,amsfonts,amsmath,amssymb,color,url,balance,graphicx,colortbl,subfigure} 
\usepackage[utf8]{inputenc}
\usepackage[english]{babel}
\usepackage{gastex} 
\usepackage[english]{algorithm2e}

\usepackage{tikz,pgfplots} 
\pgfplotsset{compat=1.8}
\usepgfplotslibrary{statistics}
\usetikzlibrary{shapes,arrows}
\definecolor{Gray}{gray}{0.95}

\newcommand{\A}{\mathcal{A}}
\newcommand{\B}{\mathcal{B}}
\newcommand{\C}{\mathcal{C}}
\newcommand{\N}{\mathcal{N}}
\newcommand{\Nat}{\mathbb{N}}
\newcommand{\X}{\mathcal{X}}
\newcommand{\E}{\mathcal{E}_S}

\newcommand{\Lk}{\mathcal{L}_{k-2}}
\newcommand{\Lkk}[1]{\mathcal{L}_{#1}}

\newcommand{\Pk}{\mathcal{P}_k}
\newcommand{\Pkk}[1]{\mathcal{P}_{#1}}

\newcommand{\Hk}{\mathcal{H}_{k}}

\newcommand{\acc}{acc-Motif}

\newcommand{\motifk}{Motifs-k(G)}

\newcommand{\algsize}{\footnotesize}

\newtheorem{problem}{Problem}

\definecolor{addcolor}{rgb}{0,.3,.7}

\newcommand{\linespreadCompact}{\linespread{1} \algsize

}
\newcommand{\linespreadDefault}{\linespread{1.5} \normalsize}

\linespreadDefault

\begin{document}

\title{An Improved Network Motif Detection Tool}
\author{Luis A. A. Meira \and Vinícius R. Máximo
\and Alvaro L. Fazenda \and Arlindo F. da Conceição}


%

\maketitle



\abstract{
Network motif provides a way to uncover the basic building blocks of most complex networks.
This task usually demands high computer processing, specially for motif with 5 or more vertices.
This paper presents an extended methodology with the following features: (i) search for motifs up to 6 vertices, (ii) multithread processing, and a (iii) new enumeration algorithm with lower complexity. 
The algorithm  to compute motifs solve isomorphism in $O(1)$ with the use of hash table.
Concurrent threads evaluates  distinct graphs.
The enumeration algorithm has smaller computational complexity.
The experiments shows better performance with respect to other methods available in literature,
allowing bioinformatic researchers to efficiently identify motifs of size 3, 4, 5, and 6.
}



\section{Introduction}

Network Motifs, or simply motifs, correspond to small patterns that recurrently appear in a complex network~\cite{barabasi2003linked}. 
They can  be considered as the basic building blocks of complex networks and their understanding may be of interest in several areas, such as Bioinformatics \cite{MotifsBio}, 
Communication \cite{MotifsWeb}, and Software Engineering \cite{MotifsSoftwareLivre}.

Finding network motifs has been a matter of attention mainly after the 2002-seminal paper from Milo \emph{et al}. \cite{MiloShen}, that proposed motifs as a way to uncover the structural design of complex networks. Nowadays, the design of efficient algorithms for network motif discovery is an up-to-date research area. Several surveys about motif detection algorithms were published in recent years \cite{ciriello2008review, Ribeiro:2009, Elisabeth2011}.


Nowadays, the main tools available to network motif search are: Fanmod~\cite{wernicke2006fanmod}, Kavosh~\cite{kashani2009kavosh}, NetMod~\cite{li2012netmode} and \acc~\cite{CBB2014}. Table~\ref{toolsTable} compares these tools, including information about the motif size and the usage of Nauty\footnote{Nauty is an algorithm to isomorphism detection~\cite{mckay2009nauty}}.
All these different methods share some features: uses direct graphs, perform a complete motif search, i.e., check for all possible isomorphic patterns and uses induced subgraphs.
It is also possible to find other tools which can search for only one isomorphic pattern~\cite{gtrie1}, or that deal with non induced subgraphs~\cite{Chen:2006}.

\begin{table*}[h]
\centering
\caption{Main features for the best well known motif search methods.}
\bigskip
\label{toolsTable}
\linespreadCompact
\begin{tabular}{|l|c|l|c|c|}
\hline
\hline 
Algorithm & Motif size  & Counting & Parallel &  Uses Nauty\\ 
\hline
\hline
FanMod & 8 & Exact / Sampling & no  & yes\\
Kavosh & 12 & Exact & no  & yes\\
NetMod  & 6  & Exact & yes  & no\\
\acc & 6 & Exact & yes  & no\\

\hline
\hline
\end{tabular}
\end{table*}

\linespreadDefault


This paper address the following problem (See~\cite{CBB2014}):

\begin{problem}[\motifk]
\label{6motif} Given a directed graph $G(V,E)$, the problem Motifs-k consists in counting  the number of connected induced subgraphs of G of size~k grouped by isomorphic distinct subgraphs of size~k. The result is an histogram 
$\Hk(G)$.
\label{problemDef}
\end{problem}

We deal  with the problem for $k$ (vertices in a subgraph) equal to 3, 4, 5, and 6.
It is important to notice the number of isomorphic distinct subgraphs of size~3,4,5, and 6 is $13$, $199$, $9,364$ and $1,530,842$, respectively. 
For $k=7$ there are $880,471,142$ distinct isomorphic patterns.


The algorithms for motif detection can be based into two main approaches: exact counting or heuristic sampling. As these names might suggest, the former approach performs a precise count of the isomorphic pattern frequency. The latter uses statistics to estimate frequency value. Several exact search-based algorithms and tools can be found in the literature, such as acc-Motif~\cite{sitis2012, CBB2014}, NetMode \cite{li2012netmode}, MAVisto~\cite{schreiber2005mavisto}, NeMoFinder~\cite{Chen:2006}, Kavosh~\cite{kashani2009kavosh} and Grochow and Kellis~\cite{Grochow:2007}. Sampling based algorithms examples are MFinder~\cite{Kashtan:2004,Mfinder2002}, Fanmod~\cite{wernicke2006fanmod} and MODA~\cite{Moda2009}.








Exact algorithms to find network motifs are generally extremely costly in terms of CPU time and memory consumption, and present restrictions on the size of motifs \cite{kashani2009kavosh}. According to Cirielo and Guerra \cite{ciriello2008review}, motif algorithms typically consist of three steps: (a) list connected subgraphs of $k$ vertices in the original graph and in a set of randomized graphs;
(b)  group them into isomorphic classes; and
(c) determine the statistical significance of the isomorphic subgraph classes by comparing their frequencies to those of an ensemble of random graphs. 
The core of this paper focus in items (a) and (b).

In 2012, we proposed optimized methods for motif detection of size 3, 4 and 5~\cite{sitis2012, CBB2014}. This paper describes an extension for \acc~including motif detection for k=6, 
 multithread and a smaller complexity in enumeration algorithm.

\textbf {Contribution:} 
We've made an algorithm for detecting motifs faster than the state of the art for motifs of size 3 up to 6.



\section{Notation and definitions}
\label{sec:defs}

This work is an extension of~\cite{accMotifs}. In this way, more attention will be given to the improvements made. The reader can refer to the previous work for more details about the algorithms.

Let $G(V,E)$ be a directed graph with $n=|V(G)|$ vertices and $m=|E(G)|$ edges. Assume that $m \geq n-1$. If $(u,v)\in E(G)$ and $(v,u)\in E(G)$, we say it is a bidirected edge. Alternatively, if only $(u,v)\in E(G)$, we say it is a directed edge. 

Given a vertex $v\in V$,  we partitioned the graph in four disjoint sets: $\A(v)$, $\B(v)$, $\C(v)$ and $\N(v)$, as follows:

$$u\in\left \{ \begin{tabular}{ll}
	$\A(v)$,&if $(u,v)\in E(G)$ and $(v,u)\in E(G)$ \\
	$\B(v)$,&if $(v,u)\in E(G)$ and $(u,v)\not\in E(G)$ \\
	$\C(v)$,&if $(u,v)\in E(G)$ and $(v,u)\not\in E(G)$ \\
    $\N(v)$,&if $(u,v)\not\in E(G)$ and $(v,u)\not\in E(G)$ \\
\end{tabular}\right .$$

It means that $\A(v)$ are the vertices with a bidirected edge to $v$.  The vertices with edges directed from $v$ are in $\B(v)$ and the vertices with edges directed to $v$ are in $\C(v)$. 
The set $\N(v)$ represent the vertices with no relationship with vertex $v$.

\setlength{\unitlength}{4mm} 
\begin{figure}[ht!]
\begin{center}
\includegraphics[width=6cm]{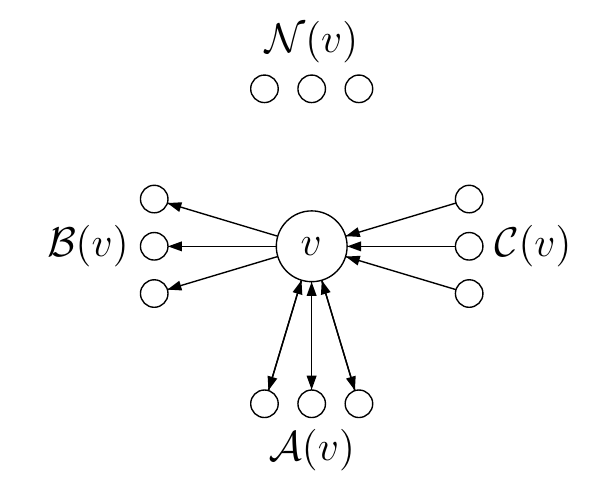}
\end{center}
\caption{Four sets definition representing a possible relationship with vertex $v$. 
}
\label{defVizOri}
\end{figure}
\setlength{\unitlength}{1mm}

Let us define $\delta(v) = \A(v) \cup \B(v) \cup \C(v)$. We define the adjacency of a set of vertices $\X \subseteq 
V$ as $adj(\X)=\left\{\cup_{v\in \X} \delta(v)\right\} \setminus \X$.  The induced graph $G[\X]$ is connected in this context.

Let $Part(adj(\X))$ be a partition of $adj(\X)$, for $|\X|\leq 4$, defined as follow.
If $\X$ is a single vertex $v\in V$, then $Part(adj(v))=\{\A(v),\B(v),C(v)\}$ as shown in Figure~\ref{defVizOri}. Note that, by definition, $\N(v)$ does not belongs to $adj(v)$.




For each pair of vertices $\{v_1,v_2\}$, the set $adj(v_1,v_2)$ is partitioned in $Part(adj(v_1,v_2))=\{AA,AB,AC,AN,BA,BB,BC,BN,CA,CB,CC,CN,NA,NB,NC\}$, where 
$AA=\A(v_1)\cap\A(v_2)$, $AB=\A(v_1)\cap\B(v_2)$ and so on. The size of $|Part(adj(v_1,v_2))|=4^2-1= 15$. Note that
$\N(v_1)\cap\N(v_2)$ is the only possible combination that does not belongs to the set $adj(v_1,v_2)$.
See Table 1 of \cite{accMotifs}.


For sets containing three vertices $\{v_1,v_2,v_3\}$, $Part(adj(v_1,v_2,v_3))=\{AAA,AAB,AAC,$ $AAN,ABA,...,NNB,NNC\}$ where $AAA = \A(v_1)\cap\A(v_2)\cap\A(v_3)$, $AAB=\A(v_1)\cap\A(v_2)\cap\B(v_3)$ and so on. The size of $|Part(adj(v_1,v_2,v_3))|=4^3-1=63$ because the set $\N(v_1)\cap\N(v_2)\cap\N(v_3)$ does not belongs to the adjacency. 

For sets $\{v_1,v_2,v_3,v_4\}$, $Part(adj(v_1,v_2,v_3,v_4))=\{AAAA,AAAB,AAAC,AAAN,$ $AABA,...,NNNB,NNNC\}$, where $AAAA = \A(v_1)\cap\A(v_2)\cap\A(v_3)\cap\A(v_4)$,  $AAAB=\A(v_1)\cap\A(v_2)\cap\A(v_3)\cap\B(v_4)$ and so on.  The size $|Part(adj(v_1,v_2,v_3))|$ is $4^4-1=255$, since $\N(v_1)\cap\N(v_2)\cap\N(v_3)\cap\N(v_4)$ dont belong to $adj(v_1,v_2,v_3)$.


Given a graph $G$, a subset of vertices $\X$, with $|\X|\leq 4$, the partition set $Part(adj(\X))$, and two sets $Y,Z\in Part(adj(\X))$.  Suppose there are no edge in $adj(\X)$. 
 The triple $\X,Y,Z$ corresponds to a motif, named  $motif(\X,Y,Z)$.
In other words, the subgraph induced $G[\X \cup y \cup z]$  corresponds to same  isomorphic graph of size $|\X|+2$ for any $y\in Y$ and $z\in Z$.  See an example (Figure~\ref{defMotif}).


\setlength{\unitlength}{4mm} 
\begin{figure}[ht!]
\begin{center}
\begin{picture}(10,10)(0,0)
\gasset{Nh=1.8,Nw=1.8,Nmr=2}
\node(0)(5,5){$v$}
\gasset{Nh=.7,Nw=.7,Nmr=2}
\node(2)(5,1){}
\nodelabel[ExtNL=y,NLangle=270,NLdist=.3](2){$\A(v)$}

\node(5)(1,5){}
\nodelabel[ExtNL=y,NLangle=180,NLdist=.3](5){$\B(v)$}



\drawedge(0,2){}
\drawedge(2,0){}
\drawedge(0,5){}
\end{picture}
\end{center}
\caption{The $motif(\X,Y,Z)$ for $\X=v$, $Y=\A(v)$ e $Z=\B(v)$.}
\label{defMotif}
\end{figure}
\setlength{\unitlength}{1mm}

Let $\Pk$ be the maximal set of distinct isomorphic connected subgraphs size $k$.
For example, $\Pkk{1}=\left\{\bullet\right\}$ has only one vertex,
$\Pkk{2}=\left\{\bullet\rightarrow \bullet,\bullet\leftrightarrow \bullet\right\}$,
$\Pkk{3}$ is a set with 13 subgraphs and so on.

\begin{figure}[ht!]
\begin{center}
\small
\begin{tabular}{cc}
\includegraphics[width=4cm]{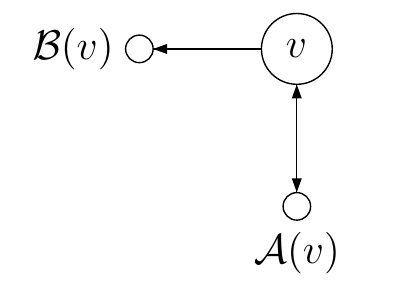}
&
\begin{tabular}{p{7cm}c|c|}
~\\
$\Lkk{1}=\{G[v_1],G[v_2],G[v_3],G[v_4]\}$\\
$\Lkk{2}=\{G[v_1,v_4],G[v_1,v_2],G[v_2,v_4],G[v_1,v_3]\}$\\
$\Lkk{3}=\{G[v_1,v_2,v_3],G[v_1,v_2,v4],G[v_1,v_3,v_4]\}$\\
~\\
Note that $G[v_2,v_3,v_4]$ is not connected thus, by definition, not belong to $\Lkk{3}$ 
\end{tabular}
\end{tabular}
\end{center}
\caption{Example of $\Lkk{r}$ for a give graph.}
\label{Lkk}
\end{figure}

The histogram $\Hk$ as described in Problem~\ref{problemDef} is a function $\Hk:\Pk\rightarrow \Nat$ associating the pattern $p\in \Pk$ to the occurrences number in $G(V,E)$.
To a graph $G(V,E)$, $\Lkk{r}(G)$ be the set of all induced connected subgraphs from $G$ with size $r$.
See an example in Figure~\ref{Lkk}.

\setlength{\unitlength}{4mm} 
\setlength{\unitlength}{1mm}


\section{Computing the Histogram $\Hk$}
\label{sec:enumeration}


The work~\cite{accMotifs} solve \motifk~using $\Lk$ as a base. The set $\Lk$ contains all connected induced  subgraphs with size $k-2$. 
The histogram $\Hk$ contains  the frequency of elements in $\Lkk{k}$ grouped by isomorphic pattern.
Thus, the algorithm to compute the histogram has complexity $\Omega(|\Lk|)$, since all  elements in $\Lk$ are considered.
Furthermore, complexity upper bounded by $o(|\Lkk{k}|)$, due the fact  the algorithm  computes the frequency for several induced subgraph in a constant time.



The diagram in Figure \ref{prog} shows the inputs and outputs in the enumeration procedure for motif of size $k$ in \acc~method.

\begin{figure}[htb!]
\vspace{-0.8cm}
\centerline{
\begin{picture}(40,25)(-35,-33)
  \node[Nadjust=wh,Nadjustdist=3,Nmr=1,Nmarks=r](A)(-15,-22){\acc}
   \node[Nh=2,Nw=2,Nmr=2,linecolor=white](B)(-33,-22){}
   \node[Nh=2,Nw=2,Nmr=2,linecolor=white](C)(2,-22){}
      \node[Nh=2,Nw=2,Nmr=2,linecolor=white](X)(-33,-19){$G(V,E)$}
            \node[Nh=2,Nw=2,Nmr=2,linecolor=white](Y)(-33,-24){$\Lk$}
                  \node[Nh=2,Nw=2,Nmr=2,linecolor=white](W)(4,-22){$\Hk$}
   \drawedge[ATnb=0,AHnb=1](B,A){}  
      \drawedge[ATnb=0,AHnb=1](A,C){}  
\end{picture}
}
\vspace{.3cm}
\caption{The \acc~algorithm. Input: $G$,~$\Lk$ and precomputed variables.
 Output:~$\Hk$. }
\label{prog}
\end{figure}

Algorithm~\ref{alg:acc} presents the motif counting process. First, it generates all induced subgraph of size $k-2$.
For $k=3$, $\Lkk{3-2}=\{G[v]~|~\forall v\in V\}$. For $k=4$, $\Lkk{4-2}=\{G[u,v]~|~\forall (u,v)\in E\}$. For $k=5$,  the set $\Lkk{5-2}$ was shown in \cite{CBB2014}. For $k=6$, the algorithm to compute 
$\Lkk{4}$
is adapted from~\cite{kashani2009kavosh}. 
The execution time to compute  $\Lkk{k-2}$  is negligible in relation to the total time.

\linespreadCompact

\begin{algorithm}[!htb]
\algsize
\LinesNumbered
\SetAlgoLined
\KwIn{Graph $G(V,E)$ and an integer $k$}
\KwOut{$\Hk$, that is the histogram for motifs of size $k$}
Compute $\Lk$. \\
\ForEach{subgraph $G[\X] \in \Lk$}
{
    Compute $Part(adj(\X))$\\
    \ForEach{$Y\in Part(adj(\X)) $}{
    \ForEach{$Z\in Part(adj(\X)) $}{
    \If{Y=Z}{
     Increment in $\Hk$ the motif $motif(\X,Y,Y)$ with  $\dbinom{|Y|} {2}$ units\\
    }\Else{
         Increment in $\Hk$ the motif $motif(\X,Y,Z)$ with  $|Y|.|Z|$ units\\
    }
    
    }    
    }
    
    \ForEach{$(u,v)\in G[ajd(\X)]$}{
    	Add one unit to motif $G[\X\cup u\cup v]$\\
		Subtract one unit from $motif(\X, Y, Z)$ such that $u\in Y\in Part(adj(\X))$ and  $v\in Z\in Part(adj(\X))$.
    }
    
}
\caption{Enumerate all subgraphs of size $k$.} 
\label{alg:acc}
\end{algorithm}

\linespreadDefault

The first diffence between the original algorithm~\cite{accMotifs} and this is that, in ~\cite{accMotifs}, each edge $(u,v)\in ajd(\X)$ increments once the variable $m_{YZ}$ for all $Y,Z\in Part(adj(\X))$.
In this solution, each edge uses $\Hk$ directly, without to create a counter $m_{YZ}$.


Note that for a constant $k$, the size $|Part(adj(\X))|$ and the number of variables $m_{xy}$ are constant, not affecting the complexity of \acc. However, the constants affect the final execution time. 

Another differcence is an improved  algorithm to compute the edges in an induced graphs as   described in Section ~\ref{sec:ind}.


\subsection{Computing Induced Subgraph $G[S]$ efficiently}
\label{sec:ind}


Given an oriented graph $G(V,E)$, with $n=|V|$ and $m=|E|$. Given a set of vertices $S$, let $\E$ be the set of edges of $G[S]$.
In this section we consider only the induced graph $G[S]$.
This section shows algorithms to obtain $\E$ efficiently. For this, it is assumed that the neighbors of $v$, $\delta(v)$, are already precomputed.


Several strategies are combined to produce an efficient algorithm to find the edges $\E$.
Let $d(v)=|\delta(v)|$ be the degree of $v$ and $D(S)=\sum_{v\in S}d(v)$ the sum of the degrees of the vertices in $S$.

\linespreadCompact

\bigskip
\hspace{-1.1cm}
\begin{minipage}{0.27\textwidth}
\begin{algorithm}[H]
\ForEach{$u\in S$}{
	\For{$v\in S$}{
 		\If{$(u,v)\in E$}{
 			$\E\gets \E\cup (u,v)$
 		}
	}
}
\caption{$\theta(|S|^2)$.}
\label{algA}
\end{algorithm}
\end{minipage}
\hspace{-.7cm}
\begin{minipage}{0.27\textwidth}
\begin{algorithm}[H]
\LinesNumbered
\SetAlgoLined
~\\
~\\
\ForEach{$u\in S$}{
	\For{$v\in \delta(u)$}{
 		\If{$v\in S$}{
 			$\E\gets \E\cup (u,v)$
 		}
	}
}
\caption{$\theta(|S|D(S))$.}
\label{algB}
\end{algorithm}
\end{minipage}
\begin{minipage}{0.27\textwidth}
\begin{algorithm}[H]
\LinesNumbered
\SetAlgoLined
~\\
~\\
~\\
\uIf{$|S|^2<|S|D(S)$}{
	\Return \textbf{Algorithm\ref{algA}}(G,S) }
	\Else{
	\Return \textbf{Algorithm\ref{algB}}(G,S)
	}    
\caption{$\theta(\min\{|S|^2,|S|D(S)\})$.}
\label{algC}
\end{algorithm}
\end{minipage}

\linespreadDefault

\bigskip


The first strategy consists of analyzing each $u,v\in S$, resulting in a complexity $|S|^2$. 
This strategy is efficient for small $|S|$. If, for example, $|S|$ is a constant, the algorithm will have complexity $O(1)$. The complexity of this algorithm regardless of whether the graph is dense or sparse.


The second strategy consists of analyzing the neighbors of $v$ for all $v\in S$. This strategy is efficient
if the graph is sparse. For example, if each vertex $v\in S$ has degree limited by a constant, the complexity will be $O(|S|)$, even for large $S$.


The third strategy choose the most efficient algorithm between Algorithm \ref{algA} and \ref{algB} according to the values of $|S|$ and $D(S)$. The resulting complexity is the minimum between the two complexity.


The fourth strategy make a partition from $S$ into $S_1$ and $S_2$. The partition is made in such a way that  vertices with smaller degree are in $S_1$ and those of greater degree are in $S_2$ (Algorithm \ref{algD}).

\linespreadCompact

\begin{algorithm}[htb!]
\LinesNumbered
\SetAlgoLined
Sort $S$ in $S'=(s'_1,\dots,s'_k)$ such that the degree is increasing.\\
Let $p$ be the value that minimizes the function $\sum_{i=1}^p d(s_i)+(|S|-p)^2$.\\
Let $S_1=(s'_1,\ldots,s'_p)$ be the $p$ vertices of smaller degree.\\
Let $S_2=(s'_{p+1},\ldots,s'_k)$ be the  $k-p$ vertices of greater degree.\\
\ForEach{$u\in S_2$}{
	\For{$v\in S_2$}{
 		\If{$(u,v)\in E$}{
 			$\E\gets \E\cup (u,v)$
 		}
	}
}
\ForEach{$u\in S_1$}{
	\For{$v\in \delta(u)$}{
 		\If{$v\in S$}{
 			$\E\gets \E\cup \{u,v\}$
 		}
	}
}


\caption{ Fourth strategy is $\Theta(\min_p\{(|S|-p)^2+\sum_{i=1}^p d(s_i)\}$ , where $s_i$ is sorted according to vertices degrees.}
\label{algD}
\end{algorithm}

\linespreadDefault


In this strategy, every pair $u,v\in S_2$ is checked, with complexity $O(|S_2|^2)$. For the lowest degree vertices, all neighbors of $d(s)$, for all $s\in S_1$, are checked. Note that the edges in $\delta(S_1, S_2)$ are computed by sweeping the neighbors of $S_1$.


Suppose that $G(V,E)$ is a sparse graph where $Hub \subset V$ such that $v \in Hub$ if and only if $d(v) \not \in O(1)$.
$Hub$ is a set of vertices whose degree is not limited to $O(1)$. Assume that $|Hub| \in O(1)$, thus the number of vertices with degree not limited to $O(1)$ is constant.
Graphs in complex networks tend to respect the above conditions. Let's call this graph class of \textit{celebrities}.


Suppose the execution of a celebrity graph by Algorithm\ref{algD}$(G,S)$.
The vertices in $Hub$ have degree greater than the vertices and $V\setminus Hub$. There is a $p$, not necessarily a minimum, such that $S_1=V \setminus Hub$ and $S_2=Hub$. For this $p$
we have that $(k-p)^2+\sum_{i=1}^p d(s_i)=|Hub|^2+\sum_{s\in V\setminus Hub}^p d(s)=\Theta(S)$.
Thus, the complexity for calculating $G[S]$ for celebrity graphs is $O(|S|)$.

\begin{table}
\linespreadCompact
\center{
\begin{tabular}{|c|p{8cm}|}
\hline
Algorithm & Complexity\\\hline
\textbf{Algorithm\ref{algA}}$(G,S)$ &  $\Theta(k^2)$ with worst case $\Theta(n^2)$\\\hline
\textbf{Algorithm\ref{algB}}$(G,S)$ &  $\Theta(D(S))$ with worst case $\Theta(m)$ 
\\\hline
\textbf{Algorithm\ref{algC}}$(G,S)$ &  $\Theta(\min\{D(S),k^2\})$ with worst case $\Theta(\min\{m,n^2\})$ \\\hline
\textbf{Algorithm\ref{algD}}$(G,S)$ &  $\Theta(k)$ \\\hline
\end{tabular}
}
\caption{Complexity of the calculation of $G[S]$ for graphs of celebrity $|S|=k$.}
\end{table}

\linespreadDefault

\subsection{Calculating Isomorphism in $O(1)$}

The use of hash to calculate isomorphisms in $ O (1) $ is not new~\cite{accMotifs,li2012netmode}. However, there are challenges to working with motifs of size $6$, due to the large number of different isomorphic patterns.

The main challenge of \acc~for $k=6$ is to compute isomorphism in $O(1)$, it is done by
pre-processing. Given a oriented graph $G(V,E)$ of size $6$, we need to compute the isomorphic pattern  representation in a hash table. 
Let $\mathcal{A}_6$ be the set of all adjacency matrix of size 6 of $G$. The size of $\mathcal{A}_6$ is  $2^{30}$ ($\approx$ 1 billion possibilities). 
A hash table was generated where the key
is an adjacency matrix $A\in \mathcal{A}_6$ and the value is a number $ID\in\{1,2,\ldots,1530842\}$ where $ID$
represents a motif. In other words, a function 
$ISO:\mathcal{A}_6\rightarrow ID$. 
After computing this table, it is possible to solve isomorphism
in $O(1)$ for any adjacency matrix.

%
%

\section{Results}
\label{resultados}

This section show the results of empirical evaluations. We compare acc-Motif with other tools present in the literature. The graphs evaluated were the same used in~\cite{accMotifs}, Table~\ref{tab:grafos} summarizes the data sets used in the experiments.

All experiments were performed using a processor IBM Power 755, of 3.3 GHz.
The first experiment consider only one thread running.


In this experiment we compared the performance of acc-Motif to Fanmode~\cite{fanmod}, Kavosh~\cite{kashani2009kavosh} and NetMode~\cite{li2012netmode} by varying $k\in\{3,4,5,6\}$ and using the graphs described in the Table \ref{tab:grafos}. In this experiment we used only one thread and the execution time was reported in milliseconds. The result of this experiment is presented in Table \ref{tab:ExpI}.

\begin{table}[!ht]
\linespreadCompact
\centering
\caption{Summary of the data sets used in the experiments.}
\label{tab:grafos}
\bigskip
\begin{tabular}{lcc}
\hline
\hline
\rowcolor{Gray}
Graph	&	$n$	&	$m$ \\
\hline
\hline
E.coli\cite{AlonDataset}	&	418	&	519\\
\rowcolor{Gray}
Levedura\cite{AlonDataset}	&	688	&	1079\\
CSphd\cite{Pajek2006}	&	1882	&	1740\\
\rowcolor{Gray}
Roget\cite{Pajek2006}	&	1022	&	5074\\
Epa\cite{Pajek2006}	&	4271	&	8965\\
\rowcolor{Gray}
California\cite{Pajek2006}	&	6175	&	16150\\
Facebook\cite{Tore}	&	1899	&	20296\\
\rowcolor{Gray}
ODLIS\cite{Pajek2006}	&	2900	&	18241\\
\hline
\hline
\end{tabular}
\end{table}

\linespreadDefault

\begin{table*}[h]
\linespreadCompact
\centering
\caption{.}
\label{tab:ExpI}
\begin{tabular}{llrrrr}
\hline
\hline 
Grafo&$k$&FanMod \cite{wernicke2006fanmod}&Kavosh \cite{kashani2009kavosh}&NetMod \cite{li2012netmode} &acc-Motif\\
\hline
\hline
\rowcolor{Gray}
E.coli&3&22.600&4.940&\textbf{0.298}&0.372\\
&4&319.139&118.416&3.525&\textbf{2.710}\\
\rowcolor{Gray}
&5&7,726.250&3,400.500&206.188&\textbf{82.547}\\
&6&164,993.500&76,107.000&5,072.250&\textbf{1,892.261}\\
\rowcolor{Gray}
Levedura&3&69.420&11.600&\textbf{0.635}&0.742\\
&4&980.733&262.050&8.208&\textbf{4.075}\\
\rowcolor{Gray}
&5&21,336.313&7,278.625&268.625&\textbf{119.602}\\
&6&408,474.250&133,385.000&8,056.500&\textbf{2,646.239}\\
\rowcolor{Gray}
CSphd&3&34.287&8.971&2.463&\textbf{1.727}\\
&4&166.792&53.475&\textbf{2.634}&3.211\\
\rowcolor{Gray}
&5&1,751.313&718.688&160.125&\textbf{43.984}\\
&6&19,705.000&8,225.500&1,102.000&\textbf{1,002.137}\\
\rowcolor{Gray}
Roget&3&164.321&24.774&6.550&\textbf{3.933}\\
&4&1,727.228&380.396&21.406&\textbf{16.509}\\
\rowcolor{Gray}
&5&32,436.375&9,524.438&510.375&\textbf{329.866}\\
&6&647,152.750&215,113.000&28,856.000&\textbf{6,450.374}\\
\rowcolor{Gray}
Epa&3&808.695&193.472&\textbf{10.155}&14.261\\
&4&41,931.931&13,008.337&390.139&\textbf{110.174}\\
\rowcolor{Gray}
&5&2,688,902.875&1,029,417.875&16,457.875&\textbf{6,239.431}\\
&6&151,426,226.500&55,472,966.500&2,548,300.750&\textbf{364,114.721}\\
\rowcolor{Gray}
California&3&1,508.350&300.445&40.313&\textbf{33.707}\\
&4&64,086.535&17,245.257&602.535&\textbf{263.317}\\
\rowcolor{Gray}
&5&4,105,529.375&1,490,712.188&29,181.625&\textbf{9,188.792}\\
&6&259,409,634.500&111,734,296.500&5,743,212.000&\textbf{566,674.477}\\
\rowcolor{Gray}
Facebook&3&3,749.735&471.141&25.506&\textbf{21.001}\\
&4&310,491.881&46,347.327&2,056.158&\textbf{597.636}\\
\rowcolor{Gray}
&5&30,622,129.500&5,300,279.938&180,014.250&\textbf{67,657.011}\\
&6&$>648 \times10^{6}$&604,958,138.500&68,980,139.250&\textbf{9,122,144.244}\\
\rowcolor{Gray}
ODLIS&3&8,674.493&848.033&44.655&\textbf{26.432}\\
&4&1,384,241.099&187,599.673&5,251.980&\textbf{774.438}\\
\rowcolor{Gray}
&5&$>162 \times10^{6}$&54,086,528.125&756,913.375&\textbf{146,503.080}\\
&6&$>648 \times10^{6}$&$>648 \times10^{6}$&454,180,868.500&\textbf{28,835,680.726}\\
\hline
\hline
\end{tabular}
\end{table*}

\linespreadDefault

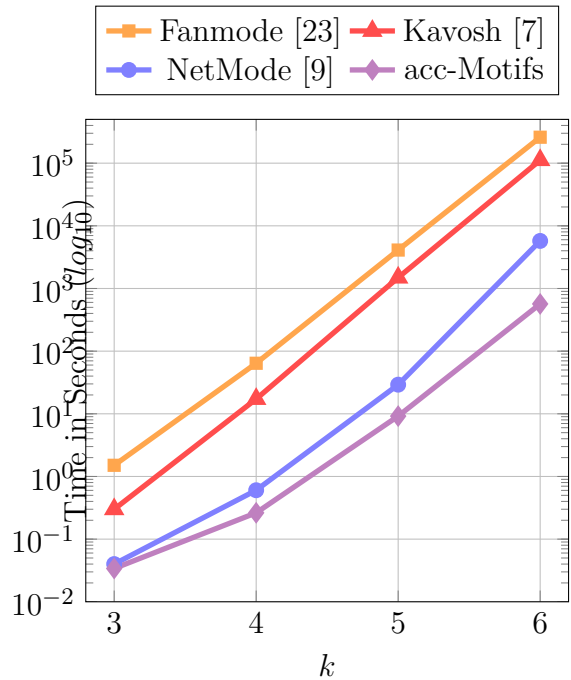
\begin{figure}[h!t]
\center
\begin{tikzpicture}
\pgfplotsset{every axis legend/.append style={
at={(0.5,1.05)},
anchor=south},
}
\begin{axis}[
width=8cm,
height=8cm,
ymode=log,
legend columns=2,
xmin=2.8,
xmax=6.2,
xtick={3,4,5,6},
ymin=0.01,
ymax=500000,
grid=major,
ylabel style={at={(0.04,0.5)}},
xlabel={$k$},
ylabel={Time in Seconds ($log_{10}$)}
]
\addplot+[
orange!70!white,
mark options={fill=orange!70!white},
line width=2pt,
mark=square*,
mark size=1.5pt
]
table[x=x,y=y] {fanmod.txt};

\addplot+[
red!70!white,
mark options={fill=red!70!white},
line width=2pt,
mark=triangle*,
mark size=2.5pt
]
table[x=x,y=y] {kavosh.txt};
\addplot+[
blue!50!white,
mark options={fill=blue!50!white},
line width=2pt,
mark=*,
mark size=2pt
]
table[x=x,y=y] {netmode.txt};

\addplot+[
violet!50!white,
mark options={fill=violet!50!white},
line width=2pt,
mark=diamond*,
mark size=2.5pt
]
table[x=x,y=y] {accMotifs.txt};

\legend{Fanmode \cite{fanmod}, Kavosh \cite{kashani2009kavosh}, NetMode \cite{li2012netmode},acc-Motifs}
\end{axis}
\end{tikzpicture}
\caption{Comparison between Fanmode, Kavosh, NetMode, \acc by varying $k\in\{3,4,5,6\}$ using the California \cite{Pajek2006} graph.}
\label{fig:variandoK}
\end{figure}

Figure \ref{fig:variandoK} shows the execution time obtained in Experiment I for the graph California. It is possible to verify that acc-Motif have performed better than the other algorithms. This result highlights the computational gain obtained by acc-Motif in relation to the best algorithms present in the literature.

According to the results presented in Table \ref{tab:ExpI}, the acc-Motif algorithm presented a performance inferior to NetMod only for instances whose computational cost is small. For larger instances, acc-Motif was superior to the other algorithms.



\begin{figure*}[h!t]
\center

\begin{tikzpicture}
\pgfplotsset{every axis legend/.append style={
at={(0.5,1.05)},
anchor=south},
}
\pgfplotsset{set layers}
\begin{axis}[
width=8cm,
height=7cm,
ymode=log,
xmode=log,
log basis x=2,
log basis y=2,
legend columns=2,
xmin=0.8,
xmax=40,
xtick={1,2,4,8,16,32,64},
ymin=128,
ymax=40000,
grid=major, 
ylabel style={at={(0.06,0.5)}},
xlabel={Number of Threads ($log_{2}$)},
ylabel={Time in Miliseconds ($log_{2}$)}
]
\addplot+[
blue!50!white,
mark options={fill=blue!50!white},
line width=2pt,
mark=*,
mark size=2pt
]
table[x=x,y=y] {netModeTime.txt};
\addplot+[
violet!50!white,
mark options={fill=violet!50!white},
line width=2pt,
mark=diamond*,
mark size=2.5pt
]
table[x=x,y=y] {accMotifTime.txt};
\legend{NetMode \cite{li2012netmode},acc-Motifs}
\end{axis}
\end{tikzpicture}
\caption{Experiment varying the number of threads for $ k = 6 $, with 511 random graphs for the graph Roget~\cite{Pajek2006}.}
\label{fig:RogetParallel}
\end{figure*}
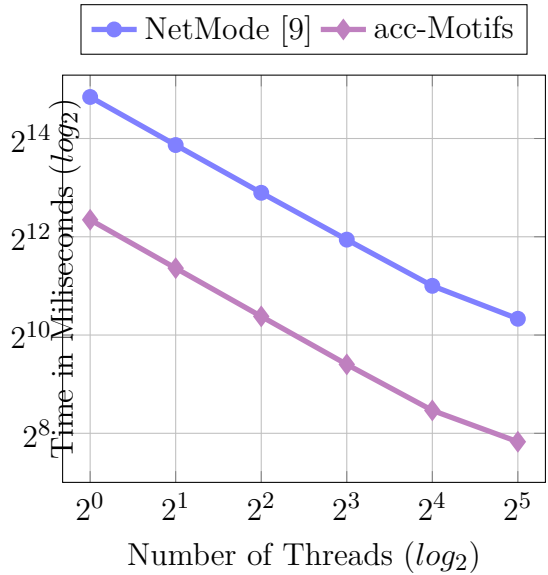

 Motifs Detection consists in enumerating the isomorphic patterns of induced subgraphs  in the original graph and in a group of random graphs. It is a paralleling problem, since each induced subgraph can be treated by a separate trhead.
 This version of \acc is multi-threaded.
 See the performance gain in tables~\ref{fig:RogetParallel} and \ref{tab:ExpII}.

\begin{table*}[h]
\centering
\linespreadCompact
\scriptsize
\caption{Experiment varying the number of threads to $ k = 6 $, with 511 random graphs.}
\label{tab:ExpII}
\begin{tabular}{llrrrrrr}
\hline
\hline 
&&\multicolumn{6}{c}{Number of Threads}\\
\cline{3-8} 
Grafo&Algoritmo&\multicolumn{1}{c}{1} &\multicolumn{1}{c}{2} &\multicolumn{1}{c}{4} &\multicolumn{1}{c}{8} &\multicolumn{1}{c}{16} &\multicolumn{1}{c}{32} \\
\hline
\hline
\rowcolor{Gray}
E,coli\cite{AlonDataset}&NetMode&4,315.15 (1)&2,171.61 (0.99)&1,120.09 (0.96)&579.34 (0.93)&299.41 (0.90)&201.00 (0.67)\\
&acc-Motif&1,008.80 (1)&509.79 (0.99)&254.90 (0.99)&134.67 (0.94)&76.86 (0.82)&52.90 (0.60)\\
\rowcolor{Gray}
Levedura\cite{AlonDataset}&NetMode&10,954.56 (1)&5,499.05 (1.00)&2,767.48 (0.99)&1,393.56 (0.98)&755.24 (0.91)&463.65 (0.74)\\
&acc-Motif&2,192.09 (1)&1,107.66 (0.99)&557.98 (0.98)&286.24 (0.96)&157.94 (0.87)&102.22 (0.67)\\
\rowcolor{Gray}
CSphd\cite{Pajek2006}&NetMode&651.88 (1)&325.96 (1.00)&169.71 (0.96)&88.89 (0.92)&49.21 (0.83)&32.75 (0.62)\\
&acc-Motif&256.34 (1)&132.00 (0.97)&76.16 (0.84)&41.79 (0.77)&28.04 (0.57)&24.44 (0.33)\\
\rowcolor{Gray}
Roget\cite{Pajek2006}&NetMode&29,396.61 (1)&14,920.73 (0.99)&7,609.44 (0.97)&3,929.97 (0.94)&2,047.88 (0.90)&1,286.86 (0.71)\\
&acc-Motif&5,198.82 (1)&2,627.17 (0.99)&1,330.58 (0.98)&676.05 (0.96)&353.37 (0.92)&226.99 (0.72)\\
\hline
\hline
\end{tabular}
\end{table*}

\linespreadDefault

\section{Conclusion}
\label{conclusion}

In this work we present a tool to detect motifs of size up to 6. Computational experiments show that \acc is the fastest tool compared to algorithms available in the literature.

We have proposed an efficient algorithm for calculating induced subgraphs.
Finally, a multi-threaded version of the program was generated.

\bibliographystyle{plain}
\bibliography{principal}

\begin{thebibliography}{10}

\bibitem{AlonDataset}
Uri Alon.
\newblock Molecular cell biology lab: Dataset.
\newblock \url{http://www.weizmann.ac.il/mcb/UriAlon/groupNetworksData.html}
  last check June 2013, 2012.

\bibitem{barabasi2003linked}
A.L. Barabasi and RE~Crandall.
\newblock Linked: The new science of networks.
\newblock {\em American journal of Physics}, 71:409, 2003.

\bibitem{Pajek2006}
Vladimir Batagelj and Andrej Mrvar.
\newblock Pajek datasets.
\newblock \url{http://vlado.fmf.uni-lj.si/pub/networks/data/}, last check June
  2013.

\bibitem{Chen:2006}
Jin Chen, Wynne Hsu, Mong~Li Lee, and See-Kiong Ng.
\newblock Nemofinder: dissecting genome-wide protein-protein interactions with
  meso-scale network motifs.
\newblock In {\em Proceedings of the 12th ACM SIGKDD international conference
  on Knowledge discovery and data mining}, KDD '06, pages 106--115, New York,
  NY, USA, 2006. ACM.

\bibitem{ciriello2008review}
G.~Ciriello and C.~Guerra.
\newblock A review on models and algorithms for motif discovery in
  protein--protein interaction networks.
\newblock {\em Briefings in functional genomics \& proteomics}, 7(2):147--156,
  2008.

\bibitem{Grochow:2007}
Joshua~A. Grochow and Manolis Kellis.
\newblock Network motif discovery using subgraph enumeration and
  symmetry-breaking.
\newblock In {\em Proceedings of the 11th annual international conference on
  Research in computational molecular biology}, RECOMB'07, pages 92--106,
  Berlin, Heidelberg, 2007. Springer-Verlag.

\bibitem{kashani2009kavosh}
Z.~Kashani, H.~Ahrabian, E.~Elahi, A.~Nowzari-Dalini, E.~Ansari, S.~Asadi,
  S.~Mohammadi, F.~Schreiber, and A.~Masoudi-Nejad.
\newblock Kavosh: a new algorithm for finding network motifs.
\newblock {\em BMC bioinformatics}, 10(1):318, 2009.

\bibitem{Kashtan:2004}
N.~Kashtan, S.~Itzkovitz, R.~Milo, and U.~Alon.
\newblock Efficient sampling algorithm for estimating subgraph concentrations
  and detecting network motifs.
\newblock {\em Bioinformatics}, 20(11):1746--1758, July 2004.

\bibitem{li2012netmode}
Xin Li, Douglas~S Stones, Haidong Wang, Hualiang Deng, Xiaoguang Liu, and Gang
  Wang.
\newblock Netmode: Network motif detection without nauty.
\newblock {\em PloS one}, 7(12):e50093, 2012.

\bibitem{MotifsSoftwareLivre}
Zhang Lin, Qian Guanqun, and Zhang Li.
\newblock Clustering analysis of motif significance profile in software
  networks.
\newblock In {\em Proceedings of the 10th WSEAS International Conference on
  Mathematical Methods and Computational Techniques in Electrical Engineering},
  pages 145--147, Stevens Point, Wisconsin, USA, 2008. World Scientific and
  Engineering Academy and Society (WSEAS).

\bibitem{MotifsBio}
Michael Lones and Andy Tyrrell.
\newblock Regulatory motif discovery using a population clustering evolutionary
  algorithm.
\newblock {\em IEEE/ACM Transactions on Computational Biology and
  Bioinformatics}, 4:403--414, 2007.

\bibitem{mckay2009nauty}
Brendan~D McKay.
\newblock nauty user’s guide (version 2.2).
\newblock Technical report, Technical Report TR-CS-9002, Australian National
  University. Available in
  \url{http://users.cecs.anu.edu.au/~bdm/nauty/nug.pdf}, 2009.

\bibitem{CBB2014}
LAA Meira, VR~Maximo, AL~Fazenda, and AF~da~Concei{\c{c}}{\~a}o.
\newblock acc-motif: Accelerated network motif detection.
\newblock {\em {IEEE/ACM Transactions on Computational Biology and
  Bioinformatics}}, PP(99), 2014.

\bibitem{accMotifs}
Luis A.~A. Meira, Vinicius~R. M\'{a}ximo, \'{A}lvaro L.~Fazenda, and Arlindo~F.
  da~Concei\c{c}\~{a}o.
\newblock {A New Approach to Count Pattern Motifs Using Combinatorial
  Techniches}.
\newblock In {\em The 8th International Conference on Signal Image Technology
  -- Workshop on Complex Network}, November 2012.

\bibitem{sitis2012}
Luis~AA Meira, Vinicius~R Maximo, Alvaro~L Fazenda, and Arlindo~F da~Conceicao.
\newblock Accelerated motif detection using combinatorial techniques.
\newblock In {\em Signal Image Technology and Internet Based Systems (SITIS),
  2012 Eighth International Conference on}, pages 744--753. IEEE, 2012.

\bibitem{MiloShen}
R.~Milo, S.~Shen-Orr, S.~Itzkovitz, N.~Kashtan, D.~Chklovskii, and U.~Alon.
\newblock Network motifs: Simple building blocks of complex networks.
\newblock {\em Science}, 298:824--887, 2002.

\bibitem{Mfinder2002}
Kashtan N., Itzkovitz S., Milo R., and Alon U.
\newblock Network motif detection tool: mfinder tool guide.
\newblock Technical report, Department of Molecular Cell Biology and Computer
  Science and Applied Mathematics, Weizman Institute of Science, Israel, 2005.

\bibitem{Moda2009}
S.~Omidi, F.~Schreiber, and A.~Masoudi-Nejad.
\newblock Moda: An efficient algorithm for network motif discovery in
  biological networks.
\newblock {\em Genes Genet. Syst}, 84:385--395, 2009.

\bibitem{Tore}
Tore Opsahl.
\newblock Datasets tore opsahl.
\newblock ~\\\url{http://toreopsahl.com/datasets/\#usairports}, 2012.

\bibitem{gtrie1}
Pedro Ribeiro and Fernando Silva.
\newblock Efficient subgraph frequency estimation with g-tries.
\newblock In Vincent Moulton and Mona Singh, editors, {\em Algorithms in
  Bioinformatics}, volume 6293 of {\em Lecture Notes in Computer Science},
  pages 238--249. Springer Berlin Heidelberg, 2010.

\bibitem{Ribeiro:2009}
Pedro Ribeiro, Fernando Silva, and Marcus Kaiser.
\newblock Strategies for network motifs discovery.
\newblock In {\em Proceedings of the 2009 Fifth IEEE International Conference
  on e-Science}, E-SCIENCE '09, pages 80--87, Washington, DC, USA, 2009. IEEE
  Computer Society.

\bibitem{schreiber2005mavisto}
F.~Schreiber and H.~Schw{\"o}bbermeyer.
\newblock Mavisto: a tool for the exploration of network motifs.
\newblock {\em Bioinformatics}, 21(17):3572--3574, 2005.

\bibitem{fanmod}
S.Wernicke and F.Rasche.
\newblock Fanmod: a tool for fast network motif detection.
\newblock {\em Bioinformatics}, 22:1152--1153, 2006.

\bibitem{wernicke2006fanmod}
S.~Wernicke and F.~Rasche.
\newblock Fanmod: a tool for fast network motif detection.
\newblock {\em Bioinformatics}, 22(9):1152--1153, 2006.

\bibitem{Elisabeth2011}
Elisabeth Wong, Brittany Baur, Saad Quader, and Chun-Hsi Huang.
\newblock Biological network motif detection: principles and practice.
\newblock {\em Brief Bioinform}, 2011.

\bibitem{MotifsWeb}
Kai-Hsiang Yang, Kun-Yan Chiou, Hahn-Ming Lee, and Jan-Ming Ho.
\newblock Web appearance disambiguation of personal names based on network
  motif.
\newblock In {\em Proceedings of the 2006 IEEE/WIC/ACM International Conference
  on Web Intelligence}, WI '06, pages 386--389, Washington, DC, USA, 2006. IEEE
  Computer Society.

\end{thebibliography}

\end{document}